\documentclass[onecolumn,showpacs,superscriptaddress,pre,nofootinbib]{revtex4}

\usepackage{graphicx}
\usepackage{amsmath,amssymb}
\usepackage{color}
\usepackage{amsfonts}

\begin{document}

\title{Generalized distributed order diffusion equations with composite time fractional derivative}

\author{Trifce Sandev}
\email{trifce.sandev@drs.gov.mk}
\affiliation{Max Planck Institute for the Physics of Complex Systems, N\"{o}thnitzer
Strasse 38, 01187 Dresden, Germany}
\affiliation{Radiation Safety Directorate, Partizanski odredi 143, P.O. Box 22, 1020
Skopje, Macedonia} 
\author{Zivorad Tomovski}
\email{zivoradt@yahoo.com}
\affiliation{Department of Mathematics, University of Rijeka, Vukovarska 58, 51000 Rijeka, Croatia}
\affiliation{Faculty of Natural Sciences and Mathematics, Institute of Mathematics, Saints Cyril and Methodius University, 1000 Skopje, Macedonia}
\author{Bojan Crnkovic}
\email{bojan.crnkovic@uniri.hr}
\affiliation{Department of Mathematics, University of Rijeka, Radmile Matejčić 2, 51000 Rijeka, Croatia}

\date{\today}

\begin{abstract}
In this paper we investigate the solution of generalized distributed order diffusion equations with composite time fractional derivative by using the Fourier-Laplace transform method. We represent solutions in terms of infinite series in Fox $H$-functions. The fractional and second moments are derived by using Mittag-Leffler functions. We observe decelerating anomalous subdiffusion in case of two composite time fractional derivatives. Generalized uniformly distributed order diffusion equation, as a model for strong anomalous behavior, is analyzed by using Tauberian theorem. Some previously obtained results are special cases of those presented in this paper.

\bigskip

\end{abstract}


\maketitle

\section{Introduction}

The fractional differential equations have attracted much attention due to their application in different fields of science \cite{YZ1,YZ2,metzler report,hilfer book,mainardi book}. Time fractional diffusion equations have been used to model anomalous diffusion processes in complex systems. They can be introduced either by using Caputo time fractional derivative or Riemann-Liouville (R-L) time fractional derivative. By using the Caputo time fractional derivative the corresponding equation which describes an anomalous diffusion process with transport exponent $0<\mu<1$ is given by \cite{metzler report}
(see also Ref.~\cite{sandev jpa2011})
\begin{align}
\label{frac diffusion eq Caputo}
\frac{\partial^{\mu}}{\partial t^{\mu}}W(x,t)=\mathcal{D}_{\mu}\frac{\partial^2}{\partial x^2}W(x,t),
\end{align}
where $W(x,t)$ is the probability distribution function (PDF),
\begin{align}
\label{Caputo_derivative}
\frac{\partial^{\mu}}{\partial t^{\mu}}f(t)={_C}D_{t}^{\mu}f(t)=\frac{1}{\Gamma(1-\mu)}\int_0
^{t}\frac{\frac{\mathrm{d}}{\mathrm{d}t'}f(t')}{(t-t')^{\mu}}\mathrm{d}t', \quad 0<\mu<1.
\end{align}
is the Caputo fractional derivative \cite{caputo,srivastava book}, and $\mathcal{D}_{\mu}$ is the generalized diffusion coefficient. Contrary to this, the R-L fractional derivative is given by \cite{hilfer book,srivastava book}
\begin{align}
{_{RL}}D_{t}^{\mu}f(t)=\frac{1}{\Gamma(1-\mu)}\frac{\mathrm{d}}{\mathrm{d}t}\int_0^{t}\frac{f(t')}{(t-t')^{\mu}}\mathrm{d}t', \quad 0<\mu<1. \nonumber
\end{align}
The initial condition is usually taken as $W(x,0+)=\delta(x)$, and the boundary conditions are given by
$W(\pm\infty,t)=\frac{\partial}{\partial x}W(\pm\infty,t)=0$ in unbounded domain in space of Lebesgue integrable functions. These are natural boundary conditions which are used in order to have an unique solution of the equation, as well as the solution to be bounded and differentiable. Note that these conditions are also used in order the Fourier transform ($\mathcal{F}\left[f(x)\right]=\int_{-\infty}^{\infty}f(x)e^{\imath kx}\mathrm{d}x$) of the second derivative to be given by $\mathcal{F}\left[f''(x)\right]=-|k|^{2}\mathcal{F}\left[f(x)\right]$. The fractional transport equation (\ref{frac diffusion eq Caputo}) is characterized by the single exponent $\mu$, since the mean square displacement (MSD) is given by $\left\langle x^{2}(t)\right\rangle=\int_{-\infty}^{\infty}x^{2}W(x,t)\mathrm{d}x=2\mathcal{D}_{\mu}\frac{t^{\mu}}{\Gamma(1+\mu)}$ \cite{metzler report}. 

Time fractional diffusion equation with single fractional exponent can be generalized by introduction of distributed
order time fractional operators \cite{chechkin1,chechkin2,chechkin chapter}. Thus, the distributed order diffusion equation is given by \cite{chechkin1}
\begin{align}\label{distributed order diff eq}
\int_0^1\tau^{\mu-1}p(\mu)\frac{\partial^{\mu}}{\partial t^{\mu}}
W(x,t)\mathrm{d}\mu=\mathcal{D}\frac{\partial^2}{\partial x^2}W(x,t),
\end{align}
where the diffusion coefficient $\mathcal{D}$ has the conventional dimension $\mathrm{m}^2/\mathrm{s}$. Here,
$\frac{\partial^{\mu}}{\partial t^{\mu}}$ is the Caputo time fractional derivative (\ref{Caputo_derivative}) of order $0<\mu<1$, $p(\mu)$ is a weight function, i.e., dimensionless non-negative function with
\begin{align}
\int_{0}^{1}p(\mu)\mathrm{d}\mu=1,
\end{align}
and $\tau$ is a time parameter with dimension $[\tau]=\mathrm{sec}$.

Generalization of time fractional and distributed order time fractional diffusion equations can be done in the framework of the CTRW theory by introduction of generalized waiting time PDF \cite{metzler report} with a given non-negative integrable function, which appears as a memory kernel from the left hand side in the diffusion equation.

Such distributed order diffusion equations has been shown to represent useful tool for modeling decelerating anomalous diffusion, ultraslow diffusive processes and strong anomaly \cite{chechkin1,chechkin2,chechkin chapter,mainardi1,mainardi3,mainardi5,mainardi4,mainardi2,sandev fcaa2015,sandev pre,saxena physica a}. These equations have been recently shown that possess multiscaling properties \cite{sandev fcaa2015,sandev pre}. 

Further generalization can be done by introduction of space fractional derivative instead of the second space derivative, for example, space fractional Riesz-Feller derivative
of order  ${\alpha}$ and skewness ${\theta }$, which is defined as a pseudo-differential operator with a symbol ${\psi_{{\alpha }}^{{\theta}}\left(k\right)}=|k|^{\alpha}e^{\imath (\mathrm{sgn}k)\theta\pi/2}$, which is the logarithm of the characteristic function of a general L\'evy strictly stable probability density with index of stability $\alpha$ and asymmetry parameter $\theta$. Its Fourier transform is given by \cite{feller}
\begin{align}\label{Riesz-Feller}
\mathcal{F}\left[D_{{\theta }}^{{\alpha
}}f\left(x\right)\right]\left(k\right)=-\psi _{{\alpha }}^{{\theta
}}\left(k\right)\mathcal{F}\left[f\left(x\right)\right]\left(k\right).
\end{align}
If ${\theta =0}$ then (\ref{Riesz-Feller}) reduces to
\begin{align}
{\mathcal{F}\left[D_{{0}}^{{\alpha
}}f\left(x\right)\right]\left(k\right)=-|k|^{{\alpha
}}\mathcal{F}\left[f\left(x\right)\right]\left(k\right).}
\end{align}
For $0<\alpha\leq2$ and $|\theta|\leq\min\left\{\alpha,2-\alpha\right\}$, the Riesz-Feller fractional derivative is given by \cite{feller}
\begin{align}\label{RF int}
D_{{\theta }}^{{\alpha
}}f\left(x\right)=\frac{\Gamma(1+\alpha)}{\pi}\left\{\sin((\alpha+\theta)\pi/2)\int_{0}^{\infty}\frac{f(x+\xi)-f(x)}{\xi^{1+\alpha}}\mathrm{d}\xi+\sin((\alpha-\theta)\pi/2)\int_{0}^{\infty}\frac{f(x-\xi)-f(x)}{\xi^{1+\alpha}}\mathrm{d}\xi\right\}.
\end{align}
The solution of the space fractional diffusion equations with space fractional Riesz-Feller derivative corresponds to the PDF obtained from the CTRW theory for L\'{e}vy flights \cite{metzler report}.

In the present paper, we introduce a new generalization of distributed order time fractional diffusion equation (\ref{distributed order diff eq}), where instead of Caputo time fractional derivative we use Hilfer-composite time fractional derivative. 

This paper is organized as follows. In Section II we formulate the problem of generalized distributed order diffusion equation, and we consider different forms of the weight function. We analyze the PDF and second moment by using the Fourier-Laplace transform method. We observe decelerating anomalous subdiffusion in the case of two composite time fractional derivatives. For uniformly distributed order diffusion equation we apply the Tauberian theorem in order to find the behavior of the second moment in the long and short time limit. Due to the contribution of all fractional exponents $\mu$ between zero and one, we observe ultraslow diffusion and strong anomaly. In Section III we consider generalized distributed order Fokker-Planck-Smoluchowski equation with two composite time fractional derivatives, and we analyze the relaxation of modes and the case of presence of an external harmonic potential. The results are exact and represented in terms of the Mittag-Leffler and Fox $H$-functions. The Summary is given in Section IV. Appendices are added at the end of the paper.

\section{Problem formulation and solution}

Inspired by the wide application of the distributed order diffusion equation (\ref{distributed order diff eq}), we consider a generalized distributed order diffusion problem with a source term $\varphi(x,t)$
\begin{align}\label{DiffusionEq}
\int_{0}^{1}\int_{0}^{1}p\left(\mu,\nu\right){_t}D_{0+}^{\mu,\nu}W\left(x,t\right)\mathrm{d}\mu\mathrm{d}\nu=\mathcal{D}\frac{\partial^{\alpha}}{\partial |x|^{\alpha}}W\left(x,t\right)+\varphi
\left(x,t\right),
\end{align}
where $W(x,t)$ is a field variable, $\mathcal{D}$ is the diffusion coefficient, $p(\mu,\nu)$ is a non-negative weight function such that
\begin{align}\label{p norm}
\int_{0}^{1}\int_{0}^{1}p(\mu,\nu)\mathrm{d}\mu\mathrm{d}\nu=1.
\end{align}
The initial conditions will depend on the form of the weight function and the boundary conditions are given by
\begin{align}
\underset{{|x|\rightarrow \infty
}}{{\text{lim}}}W\left(x,t\right)=0.
\label{Boundary}
\end{align}
Equation (\ref{DiffusionEq}) is called {\it generalized distributed order diffusion equation}.

Here, the notation ${_t}D_{0+}^{\mu,\nu}$ represents the Hilfer-composite fractional derivative of order $0<\mu<1$ and type $0\leq\nu\leq1$ defined by \cite{hilfer book}
\begin{align}\label{hilfer}
\left({_t}D_{0+}^{\mu,\nu}f\right)(t)=\left(I_{0+}^{\nu(1-\mu)}\frac{\mathrm{d}}{\mathrm{d}t}\left(I_{0+}^{(1-\nu)(1-\mu)}f\right)\right)(t),
\quad 0\leq\nu\leq1, \quad 0<\mu<1,
\end{align}
where
\begin{align}\label{RL integral}
\left(I_{0+}^{\mu}f\right)(t)=\frac{1}{\Gamma(\mu)}\int_{0}^t
\frac{f(\tau)}{(t-\tau)^{1-\mu}}\textrm{d}\tau, \quad t>0, \quad
\Re(\mu)>0,
\end{align}
is the R-L fractional integral of order $\mu>0$ \cite{srivastava book}. Note that in case where $\nu=0$ it corresponds to the classical R-L fractional derivative and in case where $\nu=1$ to the Caputo fractional derivative.

The Laplace transform of Hilfer-composite fractional derivative is given by \cite{hilfer book}
\begin{align}\label{hilfer_laplace}
\mathcal{L}\left[{_{t}}D_{0+}^{\mu,\nu}f(t)\right]&=s^\mu\mathcal{L}\left[f(t)\right]-s^{\nu(\mu-1)}\left(I_{0+}^{(1-\nu)(1-\mu)}f\right)(0+),
\end{align}
where the initial-value term
$\left(I_{0+}^{(1-\nu)(1-\mu)}f\right)(0+)$ is evaluated in the
limit as $t\rightarrow0+$ in the space of summable Lebesgue
integrable functions 
\begin{align} L(0,\infty)=\left\{f:\|f\|_{1}=\int_{0}^{\infty}|f(t)|\mathrm{d}t<\infty\right\}.\nonumber
\end{align}

The Hilfer-composite time fractional derivative has been used by Hilfer \cite{hilfer1,hilfer2}, Sandev et al. \cite{sandev jpa2011}, and Tomovski et al. \cite{TSMD physica a2012} in the analysis of fractional diffusion equations, obtaining that such models may be used in context of glass relaxation and aquifer problems. Hilfer-composite time fractional derivative was also used by Saxena et al. \cite{saxena cnsns} and Garg et al. \cite{garg} in the theory of fractional reaction-diffusion equations, where the obtained results are presented through the Mittag-Leffler (M-L) and Fox $H$-functions, as well as by Dorrego \cite{dorrego} in the ultra-hyperbolic time fractional diffusion-wave equation. Furthermore, an operational method for solving differential equations with the Hilfer-composite fractional derivative is presented in \cite{hilfer fcaa,kim}. Other cases of application of composite time fractional derivative are considered in \cite{malik,DTS 2015,furati1,garra,tatar,gu,saxena2,saxena3}, to name but a few. From the other side, Riesz-Feller fractional derivative has been used in analysis of space-time fractional diffusion equations by Mainardi et al. \cite{MPS} and Tomovski et al. \cite{TSMD physica a2012}, where they expressed the solutions in terms of Fox $H$-function. A numerical scheme for solving fractional diffusion equation with Hilfer-composite time fractional derivative and Riesz-Feller space fractional derivative is elaborated in \cite{TSMD physica a2012}.

\subsection{Two composite time fractional derivatives}

First we consider generalized distributed order diffusion equation (\ref{DiffusionEq}) with composite time fractional derivative
\begin{align}\label{DiffusionEq two delta}
\int_{0}^{1}\int_{0}^{1}p\left(\mu,\nu\right){_t}D_{0+}^{\mu,\nu}W\left(x,t\right)\mathrm{d}\mu\mathrm{d}\nu=\mathcal{D}\frac{\partial^{\alpha}}{\partial |x|^{\alpha}}W\left(x,t\right)+\varphi
\left(x,t\right),
\end{align}
where
\begin{align}\label{DiffusionEq_p two delta}
p\left(\mu,\nu \right)=\mathit{a\delta }\left(\mu -\mu
_{{1}}\right)\delta \left(\nu -\nu _{{1}}\right)+\mathit{b\delta
}\left(\mu -\mu _{{2}}\right)\delta \left(\nu -\nu
_{{2}}\right), \quad a+b=1,
\end{align}
$\mu_1>\mu_2$, with initial and boundary conditions given by
\begin{align}
\label{InitialConditions}
I_{{0+}}^{{\left(1-\nu _{{i}}\right)\left(1-\mu
_{{i}}\right)}}W\left(x,0+\right)=\delta\left({x}\right), \quad
i=1,2,
\end{align}
\begin{align}\label{Boundary ex}
\underset{{|x|\rightarrow \infty
}}{{\text{lim}}}W\left(x,t\right)=0,
\end{align}
with $\left(1-\nu _{{1}}\right)\left(1-\mu
_{{1}}\right)=\left(1-\nu _{{2}}\right)\left(1-\mu
_{{2}}\right)$. Note that the condition (\ref{p norm}) for the weight function $p(\mu,\nu)$ is satisfied since from (\ref{DiffusionEq_p two delta}) it follows $\int_{0}^{1}\int_{0}^{1}p(\mu,\nu)\mathrm{d}\mu\mathrm{d}\nu=a+b=1$. 

Generalized equation of form (\ref{DiffusionEq two delta}) for the considered form of $p(\mu,\nu)$ (\ref{DiffusionEq_p two delta}) reduces to the problem recently considered by Saxena et al. \cite{saxena cnsns} (see Section 4 for $\omega=0$), where the solutions are represented by using the M-L functions. They also considered the case where the composite time fractional derivatives are of order $1<\mu_{1},\mu_{2}\leq2$ \cite{hilfer fcaa}. For such case additional initial conditions should be taken into consideration which appear due to the Laplace transform of the Hilfer-composite time fractional derivative of order $1<\mu\leq2$, derived by Tomovski \cite{tomovski na}. Additionally to this, here we use the Fox $H$-functions to represent the exact analytical results for the fundamental solution of the equation (\ref{DiffusionEq two delta}). Furthermore, we will calculate the fractional moments, the second moment, the norm of the fundamental solution, as well as the relaxation of modes and the case of harmonic potential.

By applying Fourier-Laplace transform to equation (\ref{DiffusionEq two delta}) and taking into consideration initial and boundary conditions, for the field variable in the Fourier-Laplace space one finds  \cite{saxena cnsns}
\begin{align}\label{solution alpha FL}
\tilde{\hat{W}}(k,s)=\frac{as^{\nu_{1}(\mu_{1}-1)}}{as^{\mu_{1}}+bs^{\mu_{2}}+\mathcal{D}|k|^{\alpha}}+
\frac{bs^{\nu_{2}(\mu_{2}-1)}}{as^{\mu_{1}}+bs^{\mu_{2}}+\mathcal{D}|k|^{\alpha}}+
\frac{\tilde{\hat{\varphi}}(k,s)}{as^{\mu_{1}}+bs^{\mu_{2}}+\mathcal{D}|k|^{\alpha}},
\end{align}
where $\tilde{\hat{\varphi}}(k,s)=\mathcal{F}\left[\mathcal{L}\left[\varphi(x,t)\right]\right]$. By using the series expansion approach \cite{podlubny}, and applying inverse Laplace transform it is obtained  \cite{saxena cnsns}
\begin{align}\label{solution alpha F}
\tilde{W}(k,t)&=t^{\mu_{1}-\nu_{1}\left(\mu_{1}-1\right)-1}\sum_{n=0}^{\infty}\left(-\frac{b}{a}\right)^{n}t^{\left(\mu_{1}-\mu_{2}\right)n}E_{\mu_{1},\left(\mu_{1}-\mu_{2}\right)n+\mu_{1}-\nu_{1}\left(\mu_{1}-1\right)}^{n+1}\left(-\frac{\mathcal{D}|k|^{\alpha}}{a}t^{\mu_{1}}\right)\nonumber\\&+\frac{b}{a}t^{\mu_{1}-\nu_{2}\left(\mu_{2}-1\right)-1}\sum_{n=0}^{\infty}\left(-\frac{b}{a}\right)^{n}t^{\left(\mu_{1}-\mu_{2}\right)n}E_{\mu_{1},\left(\mu_{1}-\mu_{2}\right)n+\mu_{1}-\nu_{2}\left(\mu_{2}-1\right)}^{n+1}\left(-\frac{\mathcal{D}|k|^{\alpha}}{a}t^{\mu_{1}}\right)\nonumber\\&+\frac{1}{a}\sum_{n=0}^{\infty}\left(-\frac{b}{a}\right)^{n}\int_{0}^{t}(t-t')^{\left(\mu_{1}-\mu_{2}\right)n+\mu_{1}-1}E_{\mu_{1},\left(\mu_{1}-\mu_{2}\right)n+\mu_{1}}^{n+1}\left(-\frac{\mathcal{D}|k|^{\alpha}}{a}(t-t')^{\mu_{1}}\right)\tilde{\varphi}(k,t')\mathrm{d}t'.
\end{align}
Note that the series in three parameter M-L functions as those presented in (\ref{solution alpha F}) are convergent \cite{paneva,sandev physica a 2011}. Next we represent the M-L functions through Fox $H$-function by using relation (\ref{HML}) and then perform inverse Fourier transform (\ref{cosine H}). Thus, for the solution one can find that it is represented in terms of infinite series in Fox $H$-functions
\begin{align}\label{solution alpha}
W(x,t)&=\sum_{n=0}^{\infty}\left(-\frac{b}{a}\right)^{n}\frac{1}{n!}\frac{t^{\left(\mu_{1}-\mu_{2}\right)n+\mu_{1}-\nu_{1}\left(\mu_{1}-1\right)-1}}{\alpha|x|}\nonumber\\&\times H_{3,3}^{2,1}\left[\left.\frac{|x|}{\left(\frac{\mathcal{D}t^{\mu_1}}{a}\right)^{1/\alpha}}\right|\begin{array}{c
l}
    \displaystyle {(1,1/\alpha),\left(\left(\mu_{1}-\mu_{2}\right)n+\mu_{1}-\nu_{1}\left(\mu_{1}-1\right),\mu_{1}/\alpha\right),(1,1/2)}\\
    \displaystyle {(1,1),(n+1,1/\alpha),(1,1/2)}
  \end{array}\right]\nonumber\\&+
  \frac{b}{a}\sum_{n=0}^{\infty}\left(-\frac{b}{a}\right)^{n}\frac{1}{n!}\frac{t^{\left(\mu_{1}-\mu_{2}\right)n+\mu_{1}-\nu_{2}\left(\mu_{2}-1\right)-1}}{\alpha|x|}\nonumber\\&\times H_{3,3}^{2,1}\left[\left.\frac{|x|}{\left(\frac{\mathcal{D}t^{\mu_1}}{a}\right)^{1/\alpha}}\right|\begin{array}{c
  l}
      \displaystyle {(1,1/\alpha),\left(\left(\mu_{1}-\mu_{2}\right)n+\mu_{1}-\nu_{2}\left(\mu_{2}-1\right),\mu_{1}/\alpha\right),(1,1/2)}\\
      \displaystyle {(1,1),(n+1,1/\alpha),(1,1/2)}
    \end{array}\right]\nonumber\\&
    +\frac{1}{a\pi}\sum_{n=0}^{\infty}\left(-\frac{b}{a}\right)^{n}\int_{0}^{\infty}\cos(kx)\left(\mathcal{E}_{0+;\mu_1,\left(\mu_1-\mu_2\right)n+\mu_1}^{-\frac{\mathcal{D}k^{\alpha}}{a};n+1}\tilde{\varphi}\right)(k,t)\mathrm{d}k,
\end{align}
where $\left(\mathcal{E}_{a+;\alpha,\beta}^{\omega;\delta}f\right)(t)$ is the generalized integral operator given by (\ref{generalized integral operator}), and $\tilde{\varphi}(k,t)=\mathcal{F}\left[\varphi(x,t)\right]$. The convergence of series in Fox $H$-functions, which appear in solution (\ref{solution alpha}), can be shown in a similar way as the proof of convergence of series in M-L functions \cite{sandev physica a 2011}, by using d'Alambert criterion (ratio test) for convergence of series. Here we note that the asymptotic behavior for the solutions of time fractional diffusion equations is investigated in the literature \cite{JK,li1,li2}.

In absence of source term $\varphi(x,t)=0$, we can calculate the fractional, i.e., $q$ moment $\left\langle |x(t)|^{q}\right\rangle=\int_{-\infty}^{\infty}|x|^{q}W(x,t)\mathrm{d}x=2\int_{0}^{\infty}x^{q}W(x,t)\mathrm{d}x$, $0<q<\alpha<2$, since the second moment for $\alpha<2$ does not exist. Thus, by employing relation (\ref{integral of H}), we find
\begin{align}\label{fractional moment alpha}
\left\langle |x(t)|^{q}\right\rangle&=\frac{2}{\alpha}\left(\frac{\mathcal{D}t^{\mu_1}}{a}\right)^{q/\alpha}t^{\mu_{1}-\nu_{1}\left(\mu_{1}-1\right)-1}\nonumber\\&\times\sum_{n=0}^{\infty}\left(-\frac{b}{a}\right)^{n}\frac{1}{n!}\frac{\Gamma(1+q)\Gamma(n+1+q/\alpha)\Gamma(-q/\alpha)t^{\left(\mu_1-\mu_2\right)n}}{\Gamma(-q/2)\Gamma\left(\left(\mu_1-\mu_2\right)n+\mu_1-\nu_1\left(\mu_1-1\right)+\mu_1 q/\alpha\right)\Gamma(1+q/2)}\nonumber\\&
+\frac{2}{\alpha}\frac{b}{a}\left(\frac{\mathcal{D}t^{\mu_1}}{a}\right)^{q/\alpha}t^{\mu_{1}-\nu_{2}\left(\mu_{2}-1\right)-1}\nonumber\\&\times\sum_{n=0}^{\infty}\left(-\frac{b}{a}\right)^{n}\frac{1}{n!}\frac{\Gamma(1+q)\Gamma(n+1+q/\alpha)\Gamma(-q/\alpha)t^{\left(\mu_1-\mu_2\right)n}}{\Gamma(-q/2)\Gamma\left(\left(\mu_1-\mu_2\right)n+\mu_1-\nu_2\left(\mu_2-1\right)+\mu_1 q/\alpha\right)\Gamma(1+q/2)}.
\end{align}

Let us use $q\rightarrow0$ in (\ref{fractional moment alpha}), i.e., we calculate $\left\langle x^{0}(t)\right\rangle=\int_{-\infty}^{\infty}W(x,t)\mathrm{d}x$. Thus, we obtain
\begin{align}\label{W norm alpha}
\left\langle x^{0}(t)\right\rangle&=t^{\mu_{1}-\nu_{1}\left(\mu_{1}-1\right)-1}\sum_{n=0}^{\infty}\left(-\frac{b}{a}\right)^{n}\frac{t^{\left(\mu_1-\mu_2\right)n}}{\Gamma\left(\left(\mu_1-\mu_2\right)n+\mu_1-\nu_1\left(\mu_1-1\right)\right)}\nonumber\\&
+\frac{b}{a}t^{\mu_{1}-\nu_{2}\left(\mu_{2}-1\right)-1}\sum_{n=0}^{\infty}\left(-\frac{b}{a}\right)^{n}\frac{t^{\left(\mu_1-\mu_2\right)n}}{\Gamma\left(\left(\mu_1-\mu_2\right)n+\mu_1-\nu_2\left(\mu_2-1\right)\right)}\nonumber\\&=t^{\mu_{1}-\nu_{1}\left(\mu_{1}-1\right)-1}E_{\mu_{1}-\mu_{2},\mu_{1}-\nu_{1}(\mu_{1}-1)}\left(-\frac{b}{a}t^{\mu_{1}-\mu_{2}}\right)+\frac{b}{a}t^{\mu_{1}-\nu_{2}\left(\mu_{2}-1\right)-1}E_{\mu_{1}-\mu_{2},\mu_{1}-\nu_{2}(\mu_{2}-1)}\left(-\frac{b}{a}t^{\mu_{1}-\mu_{2}}\right).
\end{align}
From here we conclude that the solution $W(x,t)$ is not normalized, which means that it does not represent PDF. From here one can conclude that $\left\langle x^{0}(t)\right\rangle$ has power-law decay in the long time limit. Note that $\left\langle x^{0}(t)\right\rangle$ is non-negative if and only if 
\begin{align}\label{W norm alpha constraint}
&\mu_{1}-\mu_{2}\leq\mu_{1}+\nu_{1}(1-\mu_{1}), \quad 0\leq\mu_{2}+\nu_{1}(1-\mu_{1})\\
&\mu_{1}-\mu_{2}\leq\mu_{1}+\nu_{2}(1-\mu_{2}), \quad 0\leq\mu_{2}+\nu_{2}(1-\mu_{2}),
\end{align}
where we apply the properties of the completely monotone and Bernstein functions of the Laplace transform of $\left\langle x^{0}(t)\right\rangle$ (see for example \cite{garra,saxena mathematics,Tomovski2015}).

The case with one composite time fractional derivative ($b=0$) yields the known result $\left\langle x^{0}(t)\right\rangle=\frac{t^{-(1-\nu_{1})(1-\mu_{1})}}{\Gamma\left(1-(1-\nu_{1})(1-\mu_{1})\right)}$ \cite{sandev jpa2011}. The case with $\nu_{1}=0$ gives the result $\left\langle x^{0}(t)\right\rangle\sim t^{\mu_{1}-1}$, which has a experimental support \cite{bisquert1,bisquert2}. It is shown that this behavior appears in
the decaying of the charge density in semiconductors with exponential distribution of traps, as well as power law time decay of the ion-recombination isothermal luminescence
in condensed media. Thus, semiconductors with exponential distribution of traps are considered, the number of injected free carriers decays in time as a power law, due to the
trapping (power law decay of the photoconductivity) \cite{bisquert1,orenstein}. The case with $\nu_{1}=1$, gives that the solution is normalized, as it should be \cite{metzler report}. 

The case with $\nu_{1}=\nu_{2}=1$, i.e., the case of two Caputo time fractional derivatives \cite{chechkin1}, $\left\langle x^{0}(t)\right\rangle$ reduces to
\begin{align}\label{W norm alpha nu=1}
\left\langle x^{0}(t)\right\rangle=E_{\mu_{1}-\mu_{2}}\left(-\frac{b}{a}t^{\mu_{1}-\mu_{2}}\right)+\frac{b}{a}t^{\mu_{1}-\mu_{2}}E_{\mu_{1}-\mu_{2},\mu_{1}-\mu_{2}+1}\left(-\frac{b}{a}t^{\mu_{1}-\mu_{2}}\right)=1.
\end{align}
This is expected results since the solution of distributed order diffusion equation with two Caputo time fractional derivatives represents PDF, i.e., it is normalized and non-negative \cite{chechkin1,sandev pre}.

For the special case with $\alpha=2$ we arrive to the following solution
\begin{align}\label{solution alpha2}
W(x,t)&=\sum_{n=0}^{\infty}\left(-\frac{b}{a}\right)^{n}\frac{1}{n!}\frac{t^{\left(\mu_{1}-\mu_{2}\right)n+\mu_{1}-\nu_{1}\left(\mu_{1}-1\right)-1}}{2|x|}\nonumber\\&\times H_{2,2}^{2,0}\left[\left.\frac{|x|}{\left(\frac{\mathcal{D}t^{\mu_1}}{a}\right)^{1/2}}\right|\begin{array}{c
l}
    \displaystyle {\left(\left(\mu_{1}-\mu_{2}\right)n+\mu_{1}-\nu_{1}\left(\mu_{1}-1\right),\mu_{1}/2\right),(1,1/2)}\\
    \displaystyle {(1,1),(n+1,1/2)}
  \end{array}\right]\nonumber\\&+
  \frac{b}{a}\sum_{n=0}^{\infty}\left(-\frac{b}{a}\right)^{n}\frac{1}{n!}\frac{t^{\left(\mu_{1}-\mu_{2}\right)n+\mu_{1}-\nu_{2}\left(\mu_{2}-1\right)-1}}{2|x|}\nonumber\\&\times H_{2,2}^{2,0}\left[\left.\frac{|x|}{\left(\frac{\mathcal{D}t^{\mu_1}}{a}\right)^{1/2}}\right|\begin{array}{c
  l}
      \displaystyle {\left(\left(\mu_{1}-\mu_{2}\right)n+\mu_{1}-\nu_{2}\left(\mu_{2}-1\right),\mu_{1}/2\right),(1,1/2)}\\
      \displaystyle {(1,1),(n+1,1/2)}
    \end{array}\right],
\end{align}
and the fractional moments become
\begin{align}\label{fractional moment alpha=2}
\left\langle |x(t)|^{q}\right\rangle&=\left(\frac{\mathcal{D}t^{\mu_1}}{a}\right)^{q/2}t^{\mu_{1}-\nu_{1}\left(\mu_{1}-1\right)-1}\nonumber\\&\times\sum_{n=0}^{\infty}\left(-\frac{b}{a}\right)^{n}\frac{1}{n!}\frac{\Gamma(1+q)\Gamma(n+1+q/2)t^{\left(\mu_1-\mu_2\right)n}}{\Gamma\left(\left(\mu_1-\mu_2\right)n+\mu_1-\nu_1\left(\mu_1-1\right)+\mu_1 q/2\right)\Gamma(1+q/2)}\nonumber\\&
+\frac{b}{a}\left(\frac{\mathcal{D}t^{\mu_1}}{a}\right)^{q/2}t^{\mu_{1}-\nu_{2}\left(\mu_{2}-1\right)-1}\nonumber\\&\times\sum_{n=0}^{\infty}\left(-\frac{b}{a}\right)^{n}\frac{1}{n!}\frac{\Gamma(1+q)\Gamma(n+1+q/2)t^{\left(\mu_1-\mu_2\right)n}}{\Gamma\left(\left(\mu_1-\mu_2\right)n+\mu_1-\nu_2\left(\mu_2-1\right)+\mu_1 q/2\right)\Gamma(1+q/2)}\nonumber\\&=\Gamma(1+q)\left(\frac{\mathcal{D}t^{\mu_1}}{a}\right)^{q/2}t^{\mu_{1}-\nu_{1}\left(\mu_{1}-1\right)-1}E_{\mu_1-\mu_2,\mu_1-\nu_1\left(\mu_1-1\right)+\mu_1 q/2}^{1+q/2}\left(-\frac{b}{a}t^{\mu_1-\mu_2}\right)\nonumber\\&+\Gamma(1+q)\frac{b}{a}\left(\frac{\mathcal{D}t^{\mu_1}}{a}\right)^{q/2}t^{\mu_{1}-\nu_{2}\left(\mu_{2}-1\right)-1}E_{\mu_1-\mu_2,\mu_1-\nu_2\left(\mu_2-1\right)+\mu_1 q/2}^{1+q/2}\left(-\frac{b}{a}t^{\mu_1-\mu_2}\right).
\end{align}
For the special case with two Caputo time fractional derivatives ($\nu_{1}=\nu_{2}=1$) the obtained results (\ref{solution alpha2}) and (\ref{fractional moment alpha=2}) reduce to those recently obtained in \cite{sandev pre}.

From (\ref{fractional moment alpha=2}), for the second moment $q=2$ we find
\begin{align}\label{second moment alpha2}
\left\langle x^{2}(t)\right\rangle&=\frac{2}{a}\mathcal{D}t^{2\mu_{1}+\nu_{1}-\mu_{1}\nu_{1}-1}E_{\mu_{1}-\mu_{2},2\mu_{1}+\nu_{1}-\mu_{1}\nu_{1}}^{2}\left(-\frac{b}{a}t^{\mu_{1}-\mu_{2}}\right)\nonumber\\&
+\frac{2b}{a^{2}}\mathcal{D}t^{2\mu_{1}+\nu_{2}-\mu_{2}\nu_{2}-1}E_{\mu_{1}-\mu_{2},2\mu_{1}+\nu_{2}-\mu_{2}\nu_{2}}^{2}\left(-\frac{b}{a}t^{\mu_{1}-\mu_{2}}\right).
\end{align}
Here we note that, in a same way as for $\left\langle x^{2}(t)\right\rangle$, we can show that under conditions (23) and (24) the second moment (\ref{second moment alpha2}) is non-negative. 

By using the series representation of the three parameter M-L function (\ref{ML three}), for the second moment we find the following behavior $\left\langle x^{2}(t)\right\rangle\simeq t^{2\mu_{1}+\nu_{1}-\mu_{1}\nu_{1}-1}=t^{\mu_{1}-(1-\nu_{1})(1-\mu_{1})}$ in the short time limit. For the long time limit, by using relation (\ref{ML three asymptotic}), the second moment behaves as $\left\langle x^{2}(t)\right\rangle\simeq t^{\mu_{2}-(1-\nu_{2})(1-\mu_{2})}$. Since $(1-\nu_{1})(1-\mu_{1})=(1-\nu_{2})(1-\mu_{2})$ and $\mu_{1}>\mu_{2}$, we conclude that decelerating anomalous subdiffusion appears since $\mu_{2}-(1-\nu_{2})(1-\mu_{2})<\mu_{1}-(1-\nu_{1})(1-\mu_{1})<1$, where we suppose that $\mu_{1,2}-(1-\nu_{1,2})(1-\mu_{1,2})>0$. For $\nu_{1}=\nu_{2}=1$ we recover the results for distributed order diffusion equation with two fractional exponents, where the second moment from behavior $\left\langle x^{2}(t)\right\rangle\simeq t^{\mu_{1}}$ in the short time limit turns to the behavior $\left\langle x^{2}(t)\right\rangle\simeq t^{\mu_{2}}$ in the long time limit \cite{chechkin1} (see also \cite{sandev pre}).

This result is generalization of the one obtained by Chechkin et al. \cite{chechkin1} (see also \cite{sandev pre}), which can be obtained if we set $\nu_{1}=\nu_{2}=1$, i.e.,
\begin{align}
\left\langle x^{2}(t)\right\rangle&=\frac{2}{a}\mathcal{D}t^{\mu_{1}}\left[E_{\mu_{1}-\mu_{2},\mu_{1}+1}^{2}\left(-\frac{b}{a}t^{\mu_{1}-\mu_{2}}\right)
-\left(-\frac{b}{a}t^{\mu_{1}-\mu_{2}}\right)E_{\mu_{1}-\mu_{2},2\mu_{1}-\mu_{2}+1}^{2}\left(-\frac{b}{a}t^{\mu_{1}-\mu_{2}}\right)\right]\nonumber\\&=\frac{2}{a}\mathcal{D}t^{\mu_{1}}E_{\mu_{1}-\mu_{2},\mu_{1}+1}\left(-\frac{b}{a}t^{\mu_{1}-\mu_{2}}\right).
\end{align}

Another special case of second moment (\ref{second moment alpha2}) is when $b=0$ and $a=1$, i.e., \cite{sandev jpa2011}
\begin{align}\label{second moment alpha2 b0}
\left\langle x^{2}(t)\right\rangle=2\mathcal{D}\frac{t^{\mu_{1}-(1-\nu_{1})(1-\mu_{1})}}{\Gamma\left(1+\mu_{1}-(1-\nu_{1})(1-\mu_{1})\right)},
\end{align}
as it should be.

\begin{figure}[htp]
\begin{center}
  \includegraphics[width=3.5in]{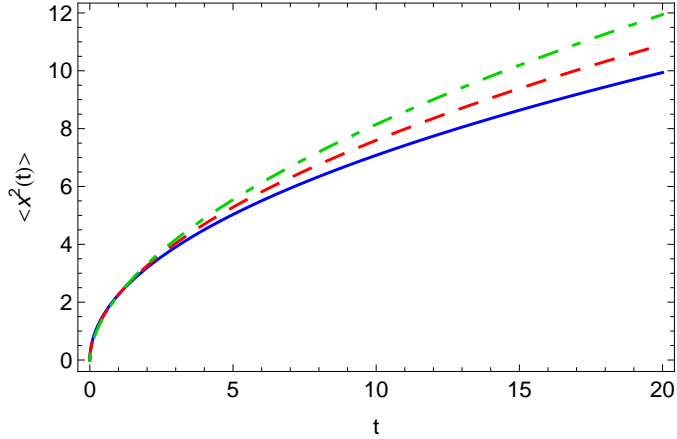}\\
  \caption{Graphical representation of second moment (\ref{second moment alpha2}) for $a=b=1/2$, $\mathcal{D}=1$, $\mu_{1}=3/4$, $\mu_{2}=5/8$; $\nu_{1}=1/4$, $\nu_{2}=1/2$ (solid blue line); $\nu_{1}=3/8$, $\nu_{2}=7/12$ (dashed red line); $\nu_{1}=1/2$, $\nu_{2}=2/3$ (dot-dashed green line).}\label{fig1a}
  \end{center}
\end{figure}

Graphical representation of the second moment (\ref{second moment alpha2}) is given in Figure \ref{fig1a}.

\subsection{Uniformly distributed order diffusion equation}

First we consider the following generalized uniformly distributed order diffusion equation
\begin{align}\label{eq alpha=2 uniform nu}
\int_{0}^{1}{_t}D_{{0+}}^{{\mu,\lambda
}}W\left(x,t\right)\mathrm{d}\mu=\mathcal{D} \frac{\partial^{2}}{\partial x^{2}}W\left(x,t\right)
\end{align} 
where the weight function in Eq.~(\ref{DiffusionEq}) is of form 
\begin{align}\label{p=1 nu}
p(\mu,\nu)=\delta(\nu-\lambda), \quad 0<\lambda<1, 
\end{align}
with initial condition
\begin{align}
\label{InitialCondition uniformly nu}
I_{0+}^{{\left(1-\lambda\right)\left(1-\mu
\right)}}W\left(x,0+\right)=\delta\left({x}\right)
\end{align}
and boundary condition (\ref{Boundary}). The weight function satisfies condition (\ref{p norm}) since $\int_{0}^{1}\int_{0}^{1}\delta(\nu-\lambda)\mathrm{d}\mu\mathrm{d}\nu=1$. Such weight function means that we take into consideration not just the influence of one fixed fractional exponent $\mu$ but the influence of all fractional exponents $\mu$ between zero and one. Due to the influence of all fractional exponents the resulting process is expected to be more anomalous than the process characterized with one single fractional exponent. This equation is a generalization of the uniformly distributed order diffusion equation where $\nu=1$ \cite{chechkin1,chechkin chapter}, which physical interpretation is that the physical process, described by uniformly distributed order diffusion equation, is lacking power-law scaling, i.e., can not be characterized by a single scaling fractional exponent $\mu$. The so-called ultraslow (or superslow) diffusion processes where the second moment has a logarithmic dependence on time are examples of processes lacking power-law scaling. The ultraslow diffusion has been observed, for example, in the Sinai model, in the motion in aperiodic environment, in iterated maps, etc (see \cite{chechkin chapter}).

From the Fourier-Laplace transform we find
\begin{align}\label{eq alpha=2 uniform nu FL}
\int_{0}^{1}\left(s^{\mu}\tilde{\hat{W}}(k,s)-s^{\lambda(\mu-1)}\right)\mathrm{d}\mu=\mathcal{D}k^{2}\tilde{\hat{W}}(k,s),
\end{align} 
from where it follows
\begin{align}\label{eq alpha=2 uniform nu FL W}
\tilde{\hat{W}}(k,s)=\frac{1}{\lambda}\frac{\frac{1-s^{-\lambda}}{\log{s}}}{\frac{s-1}{\log{s}}+\mathcal{D}k^{2}}.
\end{align}
From here, for the second moment we find
\begin{align}\label{x^2 nu}
\left\langle x^{2}(s)\right\rangle=-\left.\frac{\partial^{2}}{\partial k^{2}}\tilde{\hat{W}}(k,s)\right|_{k=0}=\frac{2\mathcal{D}}{\lambda}\frac{s^{\lambda}-1}{s^{\lambda}(s-1)^{2}}\log{s}.
\end{align}
For the the long time limit $t\rightarrow\infty$ ($s\rightarrow0$), by direct application of the Tauberian theorem (relations (\ref{tauber7}) and (\ref{tauber8}) for $\rho=\lambda$, and knowing that the logarithm function is a slowly varying function for large argument \cite{feller,mainardi book}) we can find the behavior of the second moment. Same result can be obtained as following
\begin{align}\label{x^2 nu long}
\left\langle x^{2}(t)\right\rangle&=\frac{2\mathcal{D}}{\lambda}\mathcal{L}^{-1}\left[\frac{s^{\lambda}-1}{s^{\lambda}(s-1)^{2}}\log{s}\right]\simeq\frac{2\mathcal{D}}{\lambda}\mathcal{L}^{-1}\left[s^{-\lambda}\log{\frac{1}{s}}\right]\nonumber\\&=\frac{2\mathcal{D}}{\lambda}t^{\lambda-1}\left[\log{t}/\Gamma(\lambda)-\psi(\lambda)/\Gamma(\lambda)\right]\simeq\frac{2\mathcal{D}}{\lambda}\frac{t^{\lambda-1}}{\Gamma(\lambda)}\log{t},
\end{align}
where $\psi(z)=\Gamma'(z)/\Gamma(z)$ is the polygamma function \cite{erdelyi}. From here we conclude that the particle shows more complicated behavior than the ultraslow diffusion (logarithm dependence of the second moment on time) due to the presence of the power-law term. In a similar way, for the short time limit $t\rightarrow0$ ($s\rightarrow\infty$), from the Tauberian theorem (\ref{tauber7}) and (\ref{tauber8}), where $\rho=2$, one can find the second moment. Same result can be obtained as following
\begin{align}\label{x^2 nu short}
\left\langle x^{2}(t)\right\rangle&=\frac{2\mathcal{D}}{\lambda}\mathcal{L}^{-1}\left[\frac{s^{\lambda}-1}{s^{\lambda}(s-1)^{2}}\log{s}\right]\simeq\frac{2\mathcal{D}}{\lambda}\mathcal{L}^{-1}\left[s^{-2}\log{s}\right]\nonumber\\&=\frac{2\mathcal{D}}{\lambda}t\left(1-\gamma-\log{t}\right)\simeq\frac{2\mathcal{D}}{\lambda}t\log{\frac{1}{t}},
\end{align}
where $\gamma=0.577216$ is the Euler-Mascheroni (or Euler's) constant, and we use known Laplace transform formulas with logarithm function \cite{erdelyi}. The case with $\lambda=1$ yields the well know results for uniformly distributed order diffusion equation, i.e., ultraslow diffusive behavior $\left\langle x^{2}(t)\right\rangle\simeq\log{t}$ in the long time limit, and $\left\langle x^{2}(t)\right\rangle\simeq t\log{\frac{1}{t}}$ in the short time limit \cite{chechkin1,chechkin chapter,mainardi book,sandev fcaa2015}. Therefore, the parameter $\lambda$ does not have influence on the second moment short time limit behavior. It has influence on the long time limit behavior, where strong anomaly appears.

\begin{figure}[htp]
\begin{center}
  \includegraphics[width=3.5in]{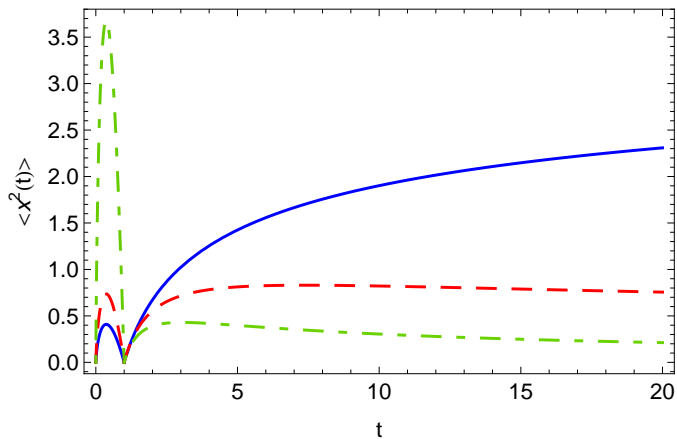}\\
  \caption{Graphical representation of second moment for short time limit (\ref{x^2 nu short}) and long time limit (\ref{x^2 nu long}) for $2\mathcal{D}=1$, $\lambda=0.9$ (solid blue line), $\lambda=0.5$ (dashed red line), $\lambda=0.1$ (dot-dashed green line).}\label{distributed}
  \end{center}
\end{figure}

Graphical representation of the second moment for the short and long time limit is given in Figure \ref{distributed}.

\section{Fokker-Planck-Smoluchowski equation with two composite time fractional derivatives}

We further consider particle dynamics confined in an external potential $V(x)$, i.e., the following generalized distributed order Fokker-Planck-Smoluchowski equation
\begin{align}\label{FPSeq}
a\,{_t}D_{0+}^{\mu_1,\nu_1}W(x,t)+b\,{_t}D_{0+}^{\mu_2,\nu_2}W(x,t)=\left[\frac{
\partial}{\partial x}\frac{V'(x)}{m\eta}+\mathcal{D}\frac{\partial^2}{
\partial x^2}\right]W(x,t),
\end{align}
with the initial and boundary conditions same as for the generalized distributed order diffusion equation (\ref{DiffusionEq two delta}). Here $m$ is the mass and $\eta$ is the friction coefficient. The separation ansatz $W(x,t)=X(x)T(t)$ yields two differential equations
\begin{align}\label{FPSeq T}
& a\,{_t}D_{0+}^{\mu_1,\nu_1}T(t)+b\,{_t}D_{0+}^{\mu_2,\nu_2}T(t)=
-\lambda T(t),\\
\label{FPSeq X}
&\left[\frac{\partial}{\partial x}\frac{V'(x)}{m\eta}+\mathcal{D}\frac{
\partial^2}{\partial x^2}\right]X(x)=-\lambda X(x),
\end{align}
where $\lambda$ is a separation constant. The solution $W(x,t)$ is given as a sum $W(x,t)=\sum_{n}^{\infty}X_{n}(x)T_{n}(t)$, where $X_{n}(x)T_{n}(t)$ are the eigenfunctions which correspond to the eigenvalues $\lambda_{n}$.

\subsection{Relaxation of modes}

From equation (\ref{FPSeq T}) by Laplace transform and taking into consideration the initial conditions, we obtain
\begin{align}\label{FPSeq T L}
T_{n}(t)=\mathcal{T}_{n}(0)\mathcal{L}^{-1}\left[a\frac{s^{\nu_1(\mu_1-1)}}{as^{\mu_1}+bs^{\mu_2}+\lambda_{n}}+b\frac{s^{\nu_2(\mu_2-1)}}{as^{\mu_1}+bs^{\mu_2}+\lambda_{n}}\right],
\end{align}
where $\mathcal{T}_{n}(0)=I_{0+}^{\left(1-\nu _{{i}}\right)\left(1-\mu
_{{i}}\right)}T_{n}(t)|_{t=0+}$. Thus, the inverse Laplace transform gives the following relaxation law
\begin{align}\label{FPSeq T sol}
\frac{T_{n}(t)}{\mathcal{T}_{n}(0)}&=\sum_{j=0}^{\infty}\left(-\frac{b}{a}\right)^{j}t^{(\mu_1-\mu_2)j+\mu_1-\nu_1(\mu_1-1)-1}E_{\mu_1,(\mu_1-\mu_2)j+\mu_1-\nu_1(\mu_1-1)}^{j+1}\left(-\frac{\lambda_{n}}{a}t^{\mu_1}\right)\nonumber\\&+\frac{b}{a}\sum_{j=0}^{\infty}\left(-\frac{b}{a}\right)^{j}t^{(\mu_1-\mu_2)j+\mu_1-\nu_2(\mu_2-1)-1}E_{\mu_1,(\mu_1-\mu_2)j+\mu_1-\nu_2(\mu_2-1)}^{j+1}\left(-\frac{\lambda_{n}}{a}t^{\mu_1}\right).
\end{align}
For the short time limit we find the following behavior
\begin{align}\label{FPSeq T sol short}
\frac{T_{n}(t)}{\mathcal{T}_{n}(0)}\simeq\frac{t^{\mu_1-\nu_1(\mu_1-1)-1}}{\Gamma\left(\mu_1-\nu_1(\mu_1-1)\right)}+\frac{b}{a}\frac{t^{\mu_1-\nu_2(\mu_2-1)-1}}{\Gamma\left(\mu_1-\nu_2(\mu_2-1)\right)},
\end{align}
where only the first terms from both series are taken. Here we note that depending on the values of parameters the second term of one of the series may be dominant in comparison to the first term from the other series. The long time limit is given by
\begin{align}\label{FPSeq T sol long}
\frac{T_{n}(t)}{\mathcal{T}_{n}(0)}&\simeq \frac{a}{b}t^{\mu_2-\nu_1(\mu_1-1)-1}E_{\mu_2,\mu_2-\nu_1(\mu_1-1)}\left(-\frac{\lambda}{b}t^{\mu_2}\right)+t^{\mu_2-\nu_2(\mu_2-1)-1}E_{\mu_2,\mu_2-\nu_2(\mu_2-1)}\left(-\frac{\lambda}{b}t^{\mu_2}\right),
\end{align}
where we apply the asymptotic expansion formula (\ref{ML three asymptotic}) and relation (\ref{ML negative alpha}) to both series.

\subsection{Harmonic potential}

The solution of equation (\ref{FPSeq X}) for the case of harmonic potential $V(x)=\frac{1}{2}m\omega^2x^2$,
where $\omega$ is a frequency, is given in terms of Hermite polynomials $H_n(z)$ (see Appendix \ref{app3})
\begin{align}
X_n(x)=\mathcal{C}_{n}H_n\left(\sqrt{\frac{m\omega^2}{2k_BT}}x\right)\exp\left(-
\frac{m\omega^2}{2k_BT}x^2\right),
\end{align}
where $\mathcal{C}_{n}$, $n=0,1,2,\dots$, are normalization constants. The eigenvalue spectrum is given by $\lambda_n=n\frac{\omega^2}{\eta_{\gamma}}$ for $n=0,1,2,\dots$. From the normalization condition
$\int_{-\infty}^{\infty}X_n^2(x)\mathrm{d}x=1$, the solution is given by
\begin{align}
W(x,t)&=\sqrt{\frac{m\omega^2}{2\pi k_BT}}\sum_n\frac{1}{2^nn!}H_n\left(\sqrt{
\frac{m\omega^2}{2k_BT}}x\right)\exp\left(-\frac{m\omega^2}{2k_BT}x^2\right)\\
&\times\left[\sum_{j=0}^{\infty}\left(-\frac{b}{a}\right)^{j}t^{(\mu_1-\mu_2)j+\mu_1-\nu_1(\mu_1-1)-1}E_{\mu_1,(\mu_1-\mu_2)j+\mu_1-\nu_1(\mu_1-1)}^{j+1}\left(-\frac{\lambda_n}{a}t^{\mu_1}\right)\right.\nonumber\\&\left.+\frac{b}{a}\sum_{j=0}^{\infty}\left(-\frac{b}{a}\right)^{j}t^{(\mu_1-\mu_2)j+\mu_1-\nu_2(\mu_2-1)-1}E_{\mu_1,(\mu_1-\mu_2)j+\mu_1-\nu_2(\mu_2-1)}^{j+1}\left(-\frac{\lambda_n}{a}t^{\mu_1}\right)\right].
\end{align}
The term $n=0$ gives
\begin{align}\label{n=0 sol}
W(x,t)&=\sqrt{\frac{m\omega^2}{2\pi k_BT}}\exp\left(-\frac{m\omega^2}{2k_BT}x^2
\right)\left[t^{\mu_1-\nu_1(\mu_1-1)-1}E_{\mu_1-\mu_2,\mu_1-\nu_1(\mu_1-1)}\left(-\frac{b}{a}t^{\mu_1-\mu_2}\right)\right.\nonumber\\&\left.+\frac{b}{a}t^{\mu_1-\nu_2(\mu_2-1)-1}E_{\mu_1-\mu_2,\mu_1-\nu_2(\mu_2-1)}\left(-\frac{b}{a}t^{\mu_1-\mu_2}\right)\right].
\end{align}
Note that in case of two Caputo time fractional derivatives ($\nu_1=\nu_2=1$), solution (\ref{n=0 sol}) represents the stationary solution $W(x,t)=\sqrt{\frac{m\omega^2}{2\pi k_BT}}\exp\left(-\frac{m\omega^2}{2k_BT}x^2
\right)$ as it should be \cite{sandev fcaa2015}.

From here one can find the first and second moments of the diffusion process governed by generalized distributed order diffusion equation in the presence of the harmonic potential $V(x)$. The first
moment follows the integro-differential equation
\begin{align}\label{diff eq first moment harmonic}
\left[a\,{_t}D_{0+}^{\mu_1,\nu_1}+b\,{_t}D_{0+}^{\mu_2,\nu_2}\right]\left\langle x(t)\right\rangle
+\frac{\omega^2}{\eta}\left\langle x(t)\right\rangle=0,
\end{align}
from where we find the relaxation law for the initial condition, given by $x_0=\int_{-\infty}^{\infty}xI_{0+}^{(1-\nu_{i})(1-\mu_{i})}W(x,0+)\mathrm{d}x=\int_{-\infty}^{\infty}x\delta(x-x_{0})\mathrm{d}x$,
\begin{align}\label{mean relaxation law}
\left\langle x(t)\right\rangle&=x_0\mathcal{L}^{-1}\left[a\frac{s^{\nu_1(\mu_1-1)}}{as^{\mu_1}+bs^{\mu_2}+\frac{\omega^{2}}{\eta}}+b\frac{s^{\nu_2(\mu_2-1)}}{as^{\mu_1}+bs^{\mu_2}+\frac{\omega^{2}}{\eta}}\right]\nonumber\\&=x_{0}\left[\sum_{j=0}^{\infty}\left(-\frac{b}{a}\right)^{j}t^{(\mu_1-\mu_2)j+\mu_1-\nu_1(\mu_1-1)-1}E_{\mu_1,(\mu_1-\mu_2)j+\mu_1-\nu_1(\mu_1-1)}^{j+1}\left(-\frac{\omega^{2}}{\eta a}t^{\mu_1}\right)\right.\nonumber\\&\left.+\frac{b}{a}\sum_{j=0}^{\infty}\left(-\frac{b}{a}\right)^{j}t^{(\mu_1-\mu_2)j+\mu_1-\nu_2(\mu_2-1)-1}E_{\mu_1,(\mu_1-\mu_2)j+\mu_1-\nu_2(\mu_2-1)}^{j+1}\left(-\frac{\omega^{2}}{\eta a}t^{\mu_1}\right)\right].
\end{align} 

For the second moment we find the following equation
\begin{align}\label{diff eq second moment harmonic}
\left[a\,{_t}D_{0+}^{\mu_1,\nu_1}+b\,{_t}D_{0+}^{\mu_2,\nu_2}\right]\left\langle x^2(t)\right\rangle+2\frac{\omega^2}{\eta}\left\langle x^2(t)\right\rangle=2\mathcal{D}\left\langle x^{0}(t)\right\rangle,
\end{align}
where $\left\langle x^{0}(t)\right\rangle$ is given by (\ref{W norm alpha}). From the Laplace transform method it follows
\begin{align}\label{x2 harmonic}
\left\langle x^2(t)\right\rangle&=x_0^2\mathcal{L}^{-1}\left[a\frac{s^{\nu_1(\mu_1-1)}}{as^{\mu_1}+bs^{\mu_2}+2\frac{\omega^{2}}{\eta}}+b\frac{s^{\nu_2(\mu_2-1)}}{as^{\mu_1}+bs^{\mu_2}+2\frac{\omega^{2}}{\eta}}\right]+\mathcal{L}^{-1}\left[
\frac{2\mathcal{D}}{as^{\mu_1}+bs^{\mu_2}+2\frac{\omega^{2}}{\eta}}\mathcal{L}\left[\left\langle x^{0}(t)\right\rangle\right]\right]\nonumber\\&=x_{0}^{2}\left[\sum_{j=0}^{\infty}\left(-\frac{b}{a}\right)^{j}t^{(\mu_1-\mu_2)j+\mu_1-\nu_1(\mu_1-1)-1}E_{\mu_1,(\mu_1-\mu_2)j+\mu_1-\nu_1(\mu_1-1)}^{j+1}\left(-2\frac{\omega^{2}}{\eta a}t^{\mu_1}\right)\right.\nonumber\\&\left.+\frac{b}{a}\sum_{j=0}^{\infty}\left(-\frac{b}{a}\right)^{j}t^{(\mu_1-\mu_2)j+\mu_1-\nu_2(\mu_2-1)-1}E_{\mu_1,(\mu_1-\mu_2)j+\mu_1-\nu_2(\mu_2-1)}^{j+1}\left(-2\frac{\omega^{2}}{\eta a}t^{\mu_1}\right)\right]\nonumber\\&+\frac{2\mathcal{D}}{a}\sum_{j=0}^{\infty}\left(-\frac{b}{a}\right)^{j}\mathcal{E}_{0+;\mu_{1},(\mu_1-\mu_2)j+\mu_1}^{-2\frac{\omega^{2}}{\eta a};j+1}\left(t^{\mu_1-\nu_1(\mu_1-1)-1}E_{\mu_1-\mu_2,\mu_1-\nu_1(\mu_1-1)}\left(-\frac{b}{a}t^{\mu_1-\mu_2}\right)\right.\nonumber\\&\left.+\frac{b}{a}t^{\mu_1-\nu_2(\mu_2-1)-1}E_{\mu_1-\mu_2,\mu_1-\nu_2(\mu_2-1)}\left(-\frac{b}{a}t^{\mu_1-\mu_2}\right)\right),
\end{align}
where $x_0^2=\int_{-\infty}^{\infty}x^{2}I_{0+}^{(1-\nu_{i})(1-\mu_{i})}W(x,0+)\mathrm{d}x=\int_{-\infty}^{\infty}x^{2}\delta(x-x_{0})\mathrm{d}x$, and the integral operator $\left(\mathcal{E}_{a+;\alpha,\beta}^{\omega;\delta}f\right)(t)$ is given by (\ref{generalized integral operator}). The case with $\nu_{1}=\nu_{2}=1$ gives the known result $\left\langle x^2(t)\right\rangle=x_0^2+\left(x_{0}^{2}-x_{th}^{2}\right)\mathcal{L}^{-1}\left[a\frac{s^{\mu_1-1}}{as^{\mu_1}+bs^{\mu_2}+2\frac{\omega^{2}}{\eta}}+b\frac{s^{\mu_2-1}}{as^{\mu_1}+bs^{\mu_2}+2\frac{\omega^{2}}{\eta}}\right]$ \cite{metzler report,sandev fcaa2015}, where $x_{0}=x(0)$, and $x_{th}^{2}=\frac{k_{B}T}{m\omega^{2}}$ is the stationary (thermal) value, which is reached in the long time limit $t\rightarrow\infty$. 

Here we note that one can analyze generalized uniformly distributed order fractional Fokker-Planck-Smoluchowski equation, which correspond to the generalized uniformly distributed order diffusion equation (\ref{eq alpha=2 uniform nu}), by applying Tauberian theorem for slowly varying functions.

\section{Summary}
We introduce generalized distributed order time fractional diffusion equation with composite time fractional derivative. The results obtained in this paper are generalization of those for time fractional and distributed order time fractional diffusion equations. We show that the solution of the equation is not normalized to one, and that it shows a power-law decay in time. The fractional moments are analyzed and the second moment is investigated in detail. Generalized uniformly distributed order diffusion equations and their second moments are analyzed by employing Tauberian theorem. Strong anomalous behavior is observed. We introduce generalized distributed order Fokker-Planck-Smoluchowski equation, and we give detail analysis of the relaxation of modes and the solution in case of external harmonic potential.

\section*{Acknowledgments}

TS acknowledges the hospitality and support from the Max-Planck Institute for the Physics of Complex Systems in Dresden, Germany. ZT is supported under the European Commission and the Croatian Ministry of Science, Education and Sports Co-Financing Agreement No. 291823. In particular, ZT acknowledges project financing from the Marie Curie FP7-PEOPLE-2011-COFUND program NEWFELPRO Grant Agreement No. 37 - Anomalous diffusion.

\appendix

\section{Mittag-Leffler and Fox $H$-functions}

For calculation of PDF and fractional moments we use three parameter Mittag-Leffler (M-L) function \cite{prabhakar}
\begin{align}\label{ML three}
E_{\alpha,\beta}^{\delta}(z)=\sum_{k=0}^{\infty}\frac{(\delta)_k}{\Gamma(\alpha
k+\beta)}\frac{z^k}{k!},
\end{align}
where $(\delta)_k=\Gamma(\delta+k)/\Gamma(\delta)$ is the Pochhammer symbol, which Laplace transform is given by \cite{prabhakar}
\begin{align}\label{ML three Laplace}
\mathcal{L}\left[t^{\beta-1}E_{\alpha,\beta}^{\delta}(\pm
at^{\alpha})\right](s)=\frac{s^{\alpha\delta-\beta}}{\left(s^{\alpha}\mp
a\right)^{\delta}}, \quad \Re(s)>|a|^{1/\alpha}.
\end{align}
The case $\delta=1$ yields the two parameter M-L function $E_{\alpha,\beta}(z)$, and the case $\beta=\delta=1$ -- one parameter M-L function $E_{\alpha}(z)$. The complete monotonicity of M-L functions is elaborated in \cite{Mainardi,Tomovski2015}.

For the three parameter M-L function the following formula \cite{sandev jmp,sandev pla,Saxena}
\begin{align}\label{ML three asymptotic}
E_{\alpha,\beta}^{\delta}(-z)=\frac{z^{-\delta}}{\Gamma(\delta)}\sum_{n=0}^{\infty}\frac{\Gamma(\delta+n)}{\Gamma(\beta-\alpha(\delta+n)
)}\frac{(-z)^{-n}}{n!}, \quad |z|>1,
\end{align}
will be used in order to analyze the asymptotic behaviors. Thus, the asymptotic expansion formula for three parameter M-L function is given by
$E_{\alpha,\beta}^{\delta}(-z)=\frac{z^{-\delta}}{\Gamma(\beta-\alpha\delta)}$
for $z\rightarrow\infty$. 

For the two parameter M-L function with negative first parameter the following relation is valid \cite{ML negative a}
\begin{align}\label{ML negative alpha}
E_{-\alpha,\beta}(z)=-z^{-1}E_{\alpha,\alpha+\beta}(1/z), \quad \alpha>0.
\end{align}

The Fox $H$-function (or simply $H$-function) is defined by the
following Mellin-Barnes integral \cite{saxena book,HMS}
\begin{align}\label{H_integral}
H_{p,q}^{m,n}\left[z\left|\begin{array}{c l}
    (a_1,A_1),...,(a_p,A_p)\\
    (b_1,B_1),...,(b_q,B_q)
  \end{array}\right.\right]=H_{p,q}^{m,n}\left[z\left|\begin{array}{c l}
    (a_p,A_p)\\
    (b_q,B_q)
  \end{array}\right.\right]=\frac{1}{2\pi\imath}\int_{\Omega}\theta(s)z^{-s}\mathrm{d}s,
\end{align}
where
$\theta(s)=\frac{\prod_{j=1}^{m}\Gamma(b_j+B_js)\prod_{j=1}^{n}\Gamma(1-a_j-A_js)}{\prod_{j=m+1}^{q}\Gamma(1-b_j-B_js)\prod_{j=n+1}^{p}\Gamma(a_j+A_js)}$,
$0\leq n\leq p$, $1\leq m\leq q$, $a_i,b_j \in C$, $A_i,B_j \in
R^{+}$, $i=1,...,p$, $j=1,...,q$. The contour $\Omega$ starting at
$c-i\infty$ and ending at $c+i\infty$ separates the poles
of the function $\Gamma(b_j+B_js)$, $j=1,...,m$ from those of the
function $\Gamma(1-a_i-A_is)$, $i=1,...,n$.

The connection between three parameter M-L function and Fox $H$-function is given by \cite{saxena book}
\begin{align}\label{HML}
E_{\alpha,\beta}^{\delta}(-z)=\frac{1}{\Gamma(\delta)}H_{1,2}^{1,1}\left[z\left|\begin{array}{l}
    (1-\delta,1)\\
    (0,1),(1-\beta,\alpha)
  \end{array}\right.\right].
\end{align}

The Fourier transform formula for Fox $H$-function is \cite{saxena book}
\begin{align}\label{cosine H}
\int_{0}^{\infty}k^{\rho-1}\cos(kx)H_{p,q}^{m,n}\left[ak^{\delta}\left|\begin{array}{c
l}
    (a_p,A_p)\\
    (b_q,B_q)
  \end{array}\right.\right]\mathrm{d}k=\frac{\pi}{x^\rho}H_{q+1,p+2}^{n+1,m}\left[\frac{x^\delta}{a}\left|\begin{array}{c
l}
     (1-b_q,B_q), (\frac{1+\rho}{2}, \frac{\delta}{2})\\
    (\rho,\delta), (1-a_p,A_p), (\frac{1+\rho}{2},\frac{\delta}{2})
  \end{array}\right.\right],
\end{align}
where $\Re\left(\rho+\delta \min_{1\leq j\leq
m}\left(\frac{b_j}{B_j}\right)\right)>1$, $x^\delta>0$,
$\Re\left(\rho+\delta \max_{1\leq j\leq
n}\left(\frac{a_j-1}{A_j}\right)\right)<\frac{3}{2}$,
$|\arg(a)|<\pi\theta/2$, $\theta>0$,
$\theta=\sum_{j=1}^{n}A_j-\sum_{j=n+1}^{p}A_j+\sum_{j=1}^{m}B_j-\sum_{j=m+1}^{q}B_j$.

The Mellin transform of the $H$-function is \cite{saxena book}
\begin{align}\label{integral of H}
&\int_{0}^{\infty}x^{\xi-1}H_{p,q}^{m,n}\left[ax\left|\begin{array}{c
l}
    (a_p,A_p)\\
    (b_q,B_q)
  \end{array}\right.\right]\mathrm{d}x=a^{-\xi}\theta(\xi),\nonumber\\
\end{align}
where $\theta(\xi)$ is defined in relation (\ref{H_integral}).

The generalized integral operator which in the kernel contains three parameter M-L function is given by \cite{prabhakar,SriTom}
\begin{align}\label{generalized integral operator}
\left(\mathcal{E}_{a+;\alpha,\beta}^{\omega;\delta}f\right)(t)=\int_{a}^{t}(t-\tau)^{\beta-1}E_{\alpha,\beta}^{\delta}\left(\omega(t-\tau)^{\alpha}\right)f(\tau)\mathrm{d}\tau,
\end{align}
where $\beta,\delta,\omega\in C$, $\Re(\alpha)>0$. The case with $\omega=0$ yields the classical R-L integral operator (\ref{RL integral}). Such generalized integral operator has been shown to have application in presentation of solution of fractional diffusion/wave equations with and without source term (external force) \cite{sandev jpa2011,sandev jpa2010,TS CMA,TS AMC,TS ND}.

\section{Tauberian theorem}

In case of distributed order diffusion equations, when the exact analytical results can not be obtained, one can analyze the asymptotic behavior of the PDFs and MSDs by applying Tauberian theorem. It states that \cite{feller} for a slowly varying function $L(t)$ at infinity, i.e., $\lim_{t\rightarrow\infty}\frac{L(at)}{L(t)}=1$, $a>0$, if
\begin{align}\label{tauber7}
\hat{R}(s)=\mathcal{L}\left[r(t)\right]\simeq s^{-\rho}L\left(\frac{1}{s}\right), \quad
s\rightarrow0, \quad \rho\geq0,
\end{align}
then
\begin{align}\label{tauber8}
r(t)=\mathcal{L}^{-1}\left[\hat{R}(s)\right]\simeq
\frac{1}{\Gamma(\rho)}t^{\rho-1}L(t), \quad t\rightarrow\infty.
\end{align}

\section{Hermite polynomials}\label{app3}

The solution of the following equation
\begin{align}\label{Hn}
y''(x)-2xy'(x)+2ny(x)=0,
\end{align}
is given in terms of the Hermite polynomials $y(x)=H_{n}(x)$, $n\in N$ \cite{erdelyi}. They are orthogonal polynomials in the range $(-\infty,\infty)$ with respect to the weighting function $e^{-x^{2}}$, i.e.,
\begin{align}\label{orthogonal}
\int_{-\infty}^{\infty}e^{-x^{2}}H_{m}(x)H_{n}(x)\mathrm{d}x=2^{n}\,n!\,\sqrt{\pi}\,\delta_{m,n}.
\end{align}

\end{document}